\documentclass[twocolumn]{aastex701}

\usepackage{amsmath}
\usepackage{txfonts}

\newcommand{\pder}[2]{\frac{\partial#1}{\partial#2}}

\usepackage[normalem]{ulem}

\definecolor{RevAddGreen}{RGB}{0, 148, 0}
\definecolor{RevDelRed}{RGB}{228, 0, 0}

\definecolor{CommentRed}{RGB}{228, 0, 0}

\begin{document}

\title{Velocity dispersion of Solar Energetic Particles in the turbulent heliosphere}

\author[orcid=0000-0002-7719-7783]{T. Laitinen}
\affiliation{Jeremiah Horrocks Institute, University of Lancashire, UK}
\email{tlmlaitinen@lancashire.ac.uk}
\author[orcid=0000-0002-7837-5780]{S. Dalla}
\affiliation{Jeremiah Horrocks Institute, University of Lancashire, UK}
\email{sdalla@lancashire.ac.uk}

\begin{abstract}
Solar Energetic Particles (SEPs) are a signature of solar eruptions, and to link them to acceleration mechanisms many studies investigate their injection time at the Sun, $t_{sun}$. We assess velocity dispersion analysis (VDA), an often-used method to derive $t_{sun}$. We use full-orbit simulations of 1--100 MeV SEP protons in our recent new model of the interplanetary magnetic turbulence superposed on a Parker Spiral magnetic field. The turbulence is described analytically as dominant transverse fluctuations that are 2D with respect to the mean field, supplemented with a minor contribution of asymptotically slab turbulence modes. We determine simulated SEP intensities for four turbulence strengths and use VDA to obtain $t_{sun}$ and the apparent path length $s$ of the SEPs, employing an SEP onset threshold to mimic a realistic energetic proton background before the SEP event. We find that turbulence strongly affects $t_{sun}$ and $s$. For weak to strong turbulence, VDA estimates of $t_{sun}$ are 2-30 minutes after the actual solar injection time, and the path lengths are 0.2-5~au longer than the Parker spiral. For very strong turbulence, the path lengths are $>5$~au, considerably longer than those typically obtained from SEP observations. We also investigate the effect of energy-dependence of the pre-event proton background, and find that different background spectra result in 5-20-minute difference in VDA injection times, depending on the heliolongitude. We conclude that in many cases VDA-derived injection times include a significant contribution from turbulence and/or the pre-event background and are not an accurate estimate of the acceleration time.
\end{abstract}

\keywords{Sun: particle emission - Sun: heliosphere - magnetic fields - turbulence - methods: numerical}

\section{Introduction} \label{sec:intro}

Charged particles are accelerated up to relativistic energies during solar eruptions. Those that escape the solar corona and propagate in the interplanetary space are observed as Solar Energetic Particles (SEPs) with in situ instruments onboard spacecraft at
various heliospheric locations. In order to understand the acceleration processes responsible for the SEPs, we must understand how they propagate through the interplanetary space from their source to the observing spacecraft. 

SEPs are guided by the interplanetary magnetic field (IMF), which on average is shaped as an Archimedean spiral, the Parker Spiral \citep{Parker1958_DynamicsInterplanetaryGas}, and is superposed by turbulent fluctuations. Due to the IMF turbulence, uncovering the origins of SEPs from the in situ observations is complicated. The transport of SEPs in the turbulent IMF has been studied for several decades \citep[see, e.g.][for a recent review]{vandenBerg2020_PrimerFocusedSolar}, with several SEP transport models introduced over time. However, the models are often computationally heavy and depend on unknown SEP transport parameters, and because of this, they are typically only used in case studies of SEP events \citep[e.g.][]{Dresing2012_LargeLongitudinalSpread, Giacalone2020_SolarEnergeticParticles, Wijsen2022_Observation-basedmodellingenergetic, Ding2022_Modeling2020November}. Particularly in large statistical observational studies \citep[e.g.][]{Vainio2013_firstSEPServerevent, Richardson2014_25MeVProton, Paassilta2017_Catalogue55-80MeV}, often simple methods of timing the SEP injection at the Sun are employed to analyse the origins of SEPs.

A popular simple method for deriving the time of SEP injection relies on the concept of velocity dispersion: the fastest SEPs arrive at the observer first. If we assume that the particles are released simultaneously, retain a constant speed during the propagation, and the first-arriving SEPs propagate without scattering and each follow the same length $s$ of path from the source to the observer, the arrival time of the first particles, often called the onset time $t_{onset}$, is given by the simple velocity dispersion equation
\begin{equation}
  \label{eq:VDA}
  t_{onset}(E) = t_{sun}+s/v
\end{equation}
where $t_{sun}$ is the time the SEPs were injected at the Sun, and $v$ is the velocity of the particles. If $t_{onset}$ and $1/v$ are known at several velocities, or energy channels, a simple linear fit will determine both $t_{sun}$ and $s$ \citep[see][on determining $t_{onset}$]{Huttunen-Heikinmaa2005_Protonheliumrelease, Zhao2019_StatisticalAnalysisInterplanetary, Posner2024_SEPClockDiscussion, Palmroos2025_NewMethodDetermining}. This Velocity Dispersion Analysis (VDA) method has been used in many studies across several decades \citep[e.g.][]{Lin1981_Energeticelectronsplasma,
  Reames1985_SolarHe-3-richevents, Torsti1998_Energetic1to,
  Krucker2000_TwoClassesSolar, Tylka2003_OnsetsReleaseTimes,
  Dalla2003_Delaysolarenergetic,
  Reames2009_SolarEnergetic-ParticleRelease,
  Vainio2013_firstSEPServerevent, Paassilta2018_Catalogue55MeV,
  Zhao2019_StatisticalAnalysisInterplanetary, Posner2024_SEPClockDiscussion} and is typically used as the first, and in many cases the only analysis method to evaluate the injection of SEPs at the Sun.

As VDA is widely used as the main tool for evaluating SEP solar injection time, it is important to understand how reliable it is. Some indication of issues in reliability are seen in the wide variety of apparent path lengths obtained with the method. \citet{Paassilta2017_Catalogue55-80MeV} analysed solar proton events at energies in the range of 1.58–-131 MeV observed by the ERNE instrument onboard the SOHO spacecraft during 1996–-2016, and found that the apparent path lengths were typically between 1 and 3~au. The path lengths were typically considerably longer than the nominal Parker Spiral length of 1.1--1.2~au, which is often assumed to be the path that the first SEPs arriving to an instrument have taken. It remains unclear whether the discrepancy between the VDA path lengths and the Parker spiral length implies also a discrepancy between the VDA-derived and the actual SEP solar injection times. 

It has been suggested that the discrepancy in the derived long path lengths, as well as solar injection time, may arise from the interplanetary propagation of SEPs. SEPs propagate through turbulent medium, which causes them to scatter as they propagate along the interplanetary magnetic field lines. 
The scattering mean free path of SEPs along the magnetic field in interplanetary space, $\lambda_\parallel$, has been estimated to be between 0.08 and 0.3~au \citep[the so-called Palmer consensus,][]{Palmer1982_Transportcoefficientslow-energy}, short compared to the typical observer location at 1~au. Further, SEP observations are typically constrained by limited statistics at the onset time, and by the pre-event background intensities either from ambient solar and galactic cosmic ray background or from preceding SEP events. Thus, it is unlikely that the first-observed SEPs have not scattered during their travel. The influence of scattering on VDA results has been investigated with SEP transport simulations, which show that particularly in events where the background is high and energy-dependent, VDA results are unreliable \citep[e.g.][]{Lintunen2004_Solarenergeticparticle, Saiz2005_EstimationSolarEnergetic, Laitinen2015_CorrectingInterplanetaryScattering}. 

It has been also proposed that the long VDA path lengths have an underlying physical cause. The random-walking of turbulent field lines has been suggested as the cause of long path lengths for SEPs
\citep[e.g.][]{Pei2006_EffectRandomMagnetic,Moradi2019_PropagationScatter-freeSolar,
  Laitinen2019_FromSunto, Chhiber2021_Magneticfieldline,
  Li2023_LagrangianStochasticModel}.
In recent work, we used a model of IMF turbulence superposed onto 
the Parker spiral magnetic field to investigate the lengths of interplanetary magnetic field lines \citep{Laitinen2023_AnalyticalModelTurbulence}. We found that
turbulence lengthens the interplanetary magnetic field lines considerably, as compared to the length of the nominal Parker spiral magnetic field. In a further study, we found that the turbulence-induced lengthening
results in a 30\%-50\% delay in arrival of the first 100-MeV protons, as compared to
scatter-free propagation along an undisturbed Parker Spiral magnetic
field \citep{Laitinen2023_Solarenergeticparticle}. The delay, and consequently the path length, was found to be strongly dependent on the relative heliolongitudinal separation between the source and the observer footpoint. However, in \citet{Laitinen2023_Solarenergeticparticle} only one SEP energy was considered and the effect of turbulence-induced field line lengthening on VDA analysis could not be studied.

The aim of this paper is to use simulations to constrain how the velocity dispersion of SEPs as detected by an observer changes with the level of turbulence in the heliosphere. This is important given how widespread the use of VDA is in the analysis of SEP events. We apply VDA to simulated 1--100 MeV proton intensity profiles derived from our test particle simulations of SEP propagation in a turbulent IMF \citep{Laitinen2023_Solarenergeticparticle}.  We investigate the effect of turbulence on both the VDA apparent path length $s$ and the obtained solar injection time $t_{sun}$. We consider four different turbulence amplitudes, corresponding to the Palmer consensus range with $\lambda_\parallel$ between 0.08 and 0.3~au \citep{Palmer1982_Transportcoefficientslow-energy}. We also investigate the effect of the pre-event energetic proton background on the VDA results by introducing an energy -dependent and energy-independent SEP event threshold. We introduce our model and parameters used in Section~\ref{sec:models}, present and discuss our results in Section~\ref{sec:results}, and draw our conclusions in Section~\ref{sec:conclusions}.

\section{Methods} \label{sec:models}

In this work, we solve the full 3D equation of motion for the SEPs \citep[e.g., Equation (3) in][]{Marsh2013_Drift-inducedPerpendicularTransport},
recording the time and heliospheric location as they pass through a spherical surface at 1~au heliocentric distance. The
heliospheric magnetic field consists of the Parker spiral superposed
by an analytic model of heliospheric turbulence, which is described in
detail in \citet{Laitinen2023_Solarenergeticparticle}. The Parker spiral is obtained by assuming a constant radial solar wind with speed 400~km/s, and solar rotation rate of $2.86533\times 10^{-6} \mathrm{rad}\; \mathrm{s}^{-1}$, resulting in a Parker spiral length of $s_{Parker}=1.158$~au between the heliocentric radial distances of SEP injection height (2 solar radii) and 1~au.

The turbulence model consists of a main 2D wave mode field, which is transverse to the
Parker spiral direction, thus reproducing the predominantly 2D structure
expected in the heliospheric plasmas 
\citep[e.g.][]{Matthaeus1990_Evidencepresencequasi-two-dimensional,
  Bieber1996_Dominanttwo-dimensionalsolar, Oughton2015_Anisotropysolarwind,Adhikari2022_2DSlabTurbulent, Adhikari2024_MHDInertialEnergycontaining}. The 2D field is combined
with a slab-like component that is asymptotically radial and azimuthal at
distances smaller and much larger than 1~au, respectively. The 2D field causes the field-lines to random-walk, resulting in the particles propagating stochastically across the Parker spiral, whereas the slab-like component causes the SEPs to scatter along the random-walking field lines \citep[e.g.][]{Jokipii1966_Cosmic-RayPropagationI, Matthaeus2003_NonlinearCollisionlessPerpendicular, Shalchi2009_NonlinearCosmicRay,vandenBerg2020_PrimerFocusedSolar}.

For the turbulence model, we use the same parametrisation as in \citet{Laitinen2025_InterplayLargescaleDrift}, with relative turbulence variance at 1~au, $dB^2/B^2$ having values 0.2, 0.6, 1 and 2 for the weak, moderate, strong and very strong turbulence cases, respectively. For 10~MeV protons, these turbulence variances correspond to standard quasilinear theory \citep{Jokipii1966_Cosmic-RayPropagationI} parallel scattering mean free paths of 0.8, 0.37, 0.22 and 0.08~au respectively, and a factor $~1.5$ smaller for 1~MeV protons. These values correspond to the range of parallel mean free paths obtained by \citep{Palmer1982_Transportcoefficientslow-energy}, the so-called Palmer consensus range. We consider the moderate turbulence case to be illustrative for typical SEP transport conditions, whereas the very strong-turbulence and weak-turbulence cases are at the the low- and high-scattering limits as suggested by the Palmer consensus. The value $dB^2/B^2=1$ is included as it is often used in SEP transport studies to represent strong turbulence.

We simulate protons at energies $E=1-100$~MeV, with a spectrum at injection proportional to $E^{-1}$. As
our model for the heliosphere currently does not include the effects of the solar wind's electric
field, the SEPs do not lose energy. 
We inject the SEPs from a region of $8^\circ$ heliolatitudinal width
at the solar equator, and across the full $360^\circ$ in heliolongitude (resulting in an 
"ensemble-average over longitude" source as in
\citet{Laitinen2023_Solarenergeticparticle}). We introduce an
ensemble-averaged heliolongitudinal coordinate
\begin{equation}    
\phi^*=\phi_{t}-\phi_{0}
\end{equation}
where $\phi_{0}$ and $\phi_{t}$ are the heliolongitudes at the
beginning of the simulation and at time $t$, respectively, for each simulated particle. Note that
the subtraction of the initial heliolongitude results in a
delta-injection in heliolongitude.

To enable investigation of a finite source size in heliolongitude, we
convolve the delta-function injection region with a kernel
$K_\phi=\left(H\{\phi-\phi_{-}\}-H\{\phi-\phi_{+}\}\right)$, where
$H\{\}$ is the Heaviside function, and $\phi_{+}=-\phi_{-}=4^\circ$, with
$\phi_{+}-\phi_{-}=8^\circ$ the width of
the source in heliolongitude.

We further consider time-extended injection, applying the often-used    
Reid-Axford injection profile
\begin{equation}
  \label{eq:Reid_Axford}
  g(t) = \frac{C}{t}\exp\left\{-\frac{\tau_a}{t}-\frac{t}{\tau_e}\right\}
\end{equation}
where $t\ge t_0\equiv 0$ is time with $t_0$ the start of the simulation, $\tau_a$ and $\tau_e$ represent the acceleration and escape
timescales, respectively, and $C$ is a normalisation constant. In this study, we use $\tau_a=1$~minute and $\tau_e=60$~minutes. With these values of $\tau_a$ and $\tau_e$, the injection reaches its maximum at 1~minute.

With the above considerations, our SEP source function is
\begin{multline}  \label{eq:inj}  
    Q_s(r, \theta, \phi, t)=\delta\{r-r_0\} \times g(t) \\
    \times \left(H\{\phi-\phi_{-}\}-H\{\phi-\phi_{+}\}\right)  \left(H\{\theta-\theta_{-}\}-H\{\theta-\theta_{+}\}\right)
\end{multline}
where $\theta_{+}=90^\circ+4^\circ$ and $\theta_{-}=90^\circ-4^\circ$.


\section{Results and Discussion}\label{sec:results}

In this study, we simulate ~100,000-~500,000 protons for 96 hours for the four turbulence cases discussed in Section~\ref{sec:models}. The SEPs are binned on 10~logarithmically equispaced bins with boundaries
$\log{E_i[\mathrm{MeV}]}=2 i/10$. We refer to these bins as energy
channels with the representative energy determined as the arithmetic mean
of the boundaries of each bin, following the convention typically used
in SEP instrument data. The simulation data is further binned at 1~minute temporal resolution.

We introduce fictional observers located at 1~au and different heliolongitudes at $20^\circ$ intervals. We present the observer locations in terms of relative heliolongitude, as
\begin{equation}\label{eq:deltaphi}
    \Delta\phi=\phi_{src}-\phi^*_{fpt}
\end{equation}
where $\phi_{src}=0^\circ$ is the centre of the source region in
heliolongitude, and $\phi^*_{fpt}$ is the heliolongitude of the
footpoint that connects the observer to the solar surface
along the nominal Parker spiral.

We bin the counts of simulated particles in $20^\circ$ heliolongitudinal bins. We then derive synthetic intensities $I(E, t, \Delta\phi)$ as function of time at observer location $\Delta\phi$ and energy channel $E$ and determine SEP 1-au onset times as a function of energy. Using these onset times, we apply the VDA analysis to derive $t_{sun}$ and $s$ by fitting the onset times and particle velocities to Equation~(\ref{eq:VDA}).

In order to carry out the VDA analysis, we need to specify a criterion for determining the onset time at 1~au, $t_{onset}(E, \Delta\phi)$. In this work, we define $t_{onset}(E, \Delta\phi)$ as the time when the proton intensity at the given heliolongitude increases above a given threshold $I_{onset}(E,\Delta\phi)$ given by
    \begin{equation} \label{eq:local}
        I_{onset}(E,\Delta\phi)=f(E) \max\{I(E, t, \Delta\phi)\}
    \end{equation}
where $f(E)$ is a specified threshold fraction of the maximum intensity at that observer location, $\max\{I(E, t, \Delta\phi)\}$. $I_{onset}(E,\Delta\phi)$ can be taken to represent the pre-event background intensities. This definition of the threshold mimics the real-world SEP analysis situation, where the onset time can only be determined when the SEP intensities clearly exceed the pre-event intensities, often affected by solar energetic particle background, possible previous SEP events, and instrumental effects.

The fraction $f(E)$ of the maximum intensity used in Equation~(\ref{eq:local}) to define the threshold is in general energy-dependent since for observed SEP events the peak spectrum is likely not the same as the pre-event background spectrum. The low-energy (0.1-10 MeV) proton background spectrum can be characterised as a power-law spectrum $E^{-\gamma}$ with $\gamma\approx 3$ during solar minimum, with $\gamma$ slightly smaller during more active portions of the solar cycle \citep{Logachev2002_EnergySpectraLowFlux}. The spectrum turns to $\propto E^{1}$ due to galactic cosmic rays at a few (at activity minimum) to 20-30~MeV (at activity maximum), thus affecting the overall background spectrum considerably in the energy range of 1-100~MeV. The SEP event spectra, on the other hand, vary considerably from event to event. The low-energy spectrum is often hard, with $\gamma\sim 1$, however the hard power law is broken to a softer $\gamma\sim 4$ at energies 2-50 MeV \citep{Mewaldt2005_ProtonHeliumElectrona,Mewaldt2012_EnergySpectraCompositiona}
Thus, the SEP event spectrum in the range of 1-100~MeV is typically steeper than the background spectrum. Based on these considerations, we consider here a simple energy-dependence of $f(E)$ as 
\begin{equation}\label{eq:thresh_E}
    f(E)=f_0 \left(\frac{E}{E_0}\right)^{-1}
\end{equation}
where $E_0=100\,\mathrm{MeV}$ and $f_0=0.1$. Thus, in our work the SEP event onset threshold is 10\% of the peak intensity at 100~MeV, and 0.1\% of the peak intensity at 1~MeV. We note that the choice of the level of the threshold has a significant effect on VDA results, as shown in  \citet{Lintunen2004_Solarenergeticparticle, Saiz2005_EstimationSolarEnergetic, Laitinen2010_EffectPre-eventBackground, Laitinen2015_CorrectingInterplanetaryScattering}.

Finally, we
require that at least 5 simulated particles are recorded in the binned simulation data at onset. We also require that the onset time can be determined in at
least 4~energy channels, following the above criteria.

\begin{figure}
    \centering
    \includegraphics[width=\linewidth]{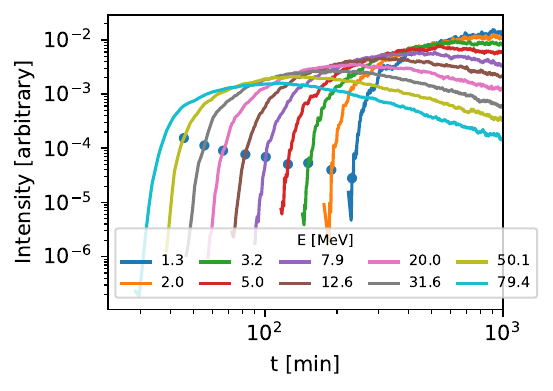}
    \caption{Intensity time profiles for the simulated SEP event at $\Delta\phi=0^\circ$ (curves) for the moderate turbulence case, along with the onset times (blue circles) obtained with the criteria outlined in the text.}
    \label{fig:timeprof}
\end{figure}

\begin{figure}
    \centering
    \includegraphics{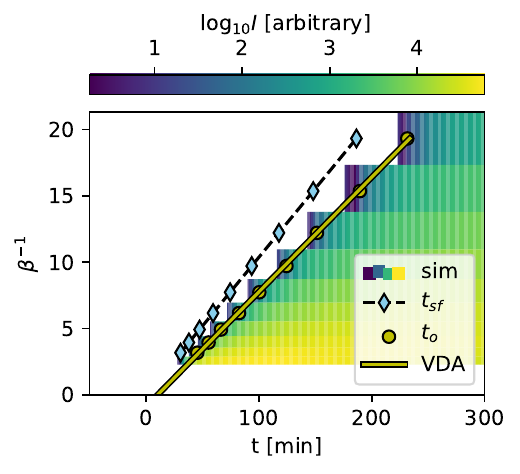}
    \caption{The contour shows the simulated proton intensities as a function of inverse of velocity (in light-speed units) and time at $\Delta\phi=0^\circ$, for the moderate turbulence case. The yellow symbols the onset times determined with  Equation~(\ref{eq:local}), and the yellow curve the fit of the onset times to Equation~(\ref{eq:VDA}). The blue diamonds, connected with the dashed line, show the arrival time of scatter-free protons along nominal Parker spiral.}\label{fig:veldisp}
  \end{figure}

In Figure~\ref{fig:timeprof}, we show the simulated time-intensity curves at a fictional observer at $\Delta\phi=0^\circ$ for the moderate turbulence case. The onset times, determined according to our criteria, are shown with the blue filled circles intersecting each time-intensity curve.
In Fig.~\ref{fig:veldisp},  we show a contour of the proton intensity (multiplied with the energy channel width) as a function of time and
  inverse of velocity, $\beta^{-1}=c/v$, where $c$ is the light speed, again at $\Delta\phi=0^\circ$ for the moderate turbulence case. We can see a clear linear velocity dispersion in the onset times, determined by Equation~(\ref{eq:local}) (yellow circles). The blue diamonds, connected with dashed line, show the arrival times expected for scatter-free propagation along an undisturbed Parker spiral within the energy channels, $t_{sf}=s_{Parker}/v$. As can
  be seen, the observed onset times are larger than the scatter-free times, $t_{onset}>t_{sf}$ for all energy channels. The solid yellow line in Figure~\ref{fig:veldisp} shows the VDA fit. The fitting procedure for the moderate turbulence case at $\Delta\phi=20^\circ$ resulted in the solar injection time of $t_{sun}=10$~minutes, relative to the simulation start at t=0. The apparent path length from the fit was found to be $s=1.38$~au, 20\% longer than $s_{Parker}$. 

\begin{figure}
    \centering
    \includegraphics{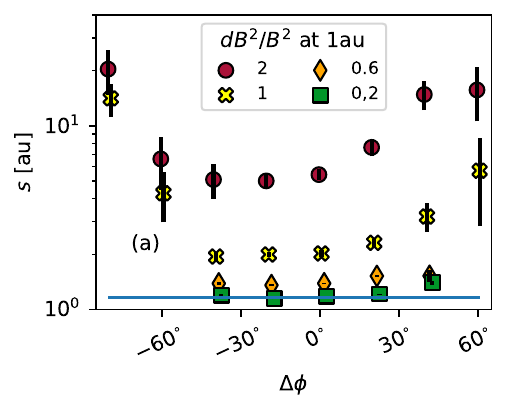}
    \includegraphics{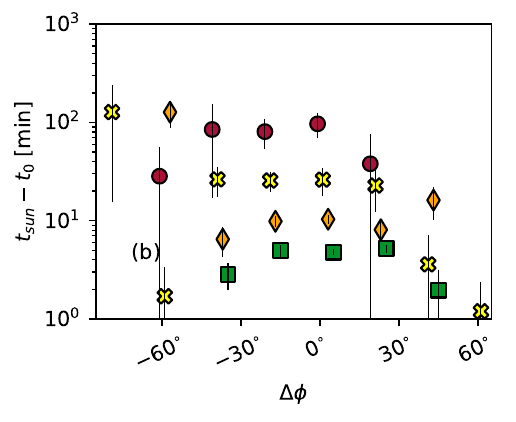}
    \caption{The VDA apparent path length (a) and solar injection time relative to the simulation injection time $t_0$ (b) of SEPs as a function of the observer heliolongitude for the very strong (red circle),strong (yellow cross), moderate (orange diamond) and weak (green square) turbulence and energy-dependent onset threshold fraction $f(E)$. The solid blue line shows the length of the nominal Parker Spiral. In panel (b), the solar injection times for $\Delta\phi=-80^\circ$, $40^\circ$ and $60^\circ$ for the high turbulence case are omitted for clarity, as they are at $t_{sun}<-100$ minutes. Note that the points from different turbulence sets are staggered by $1^\circ$ to avoid overlapping of error bars.}\label{fig:pathlen_time}
\end{figure}

In Figure~\ref{fig:pathlen_time} we show the results of the VDA
fitting for a number of fictional observers at different relative heliolongitudes $\Delta\phi$ at 1~au. In panel~(a) , we show the apparent path length $s$ with the red circles, yellow crosses, orange diamonds and olive squares representing the path length for very strong, strong, moderate and weak turbulence, respectively. The blue horizontal line shows the length of the nominal Parker spiral $s_{Parker}$. In Figure~\ref{fig:pathlen_time}~(b) we show the VDA solar injection time $t_{sun}$ of the protons, as a function of $\Delta\phi$, with the same markers and colours as in Figure~\ref{fig:pathlen_time}~(a) . Note that we omit the solar release time for $\Delta\phi=-80^\circ$, $40^\circ$ and $60^\circ$ for the high turbulence for clarity as they are at $t_{sun}<-100$ minutes.

As we can see in Figure~\ref{fig:pathlen_time}, the turbulence amplitude affects the heliolongitudinal extent of the analysed SEP event, the path length and the obtained solar release time $t_{sun}$ significantly. For the weak-turbulence case, $dB^2/B^2=0.2$ at 1~au, the VDA solar injection time $t_{sun}$ is within 10 minutes from the beginning of the injection of the simulated SEPs. The apparent path length $s$ is of the order of the nominal distance along the Parker field line, $s_{Parker}$, with slightly longer path lengths at positive $\Delta\phi$. The good accuracy of VDA in this case can be attributed to the reduced scattering and field line meandering in the low-amplitude turbulence, resulting in almost scatter-free propagation of the first SEPs along an almost-undisturbed Parker spiral. We note that the first SEPs are not completely scatter-free: the mean pitch angle cosine of the protons arriving at 1~au before the onset times $t_{onset}(E)$ was found to be $\overline{\mu}=0.8$. This clearly shows that the VDA assumption of scatter-free first SEPs fails even in the case of weak turbulence.

In the moderate turbulence case, with $dB^2/B^2=0.6$ at 1~au, the SEPs have access to similar range of heliolongitudes. The VDA solar injection time $t_{sun}$ ranges between 5-15 minutes later than the actual solar injection time. The VDA apparent path length for the moderate turbulence are about 10-25\% longer than the Parker spiral length near the best connection, thus longer than for the weak turbulence case, with positive $\Delta\phi$ showing larger path lengths. The trend of increasing apparent path length and delay continues for the strong turbulence case,  with $dB^2/B^2=1$ at 1~au, for which the $t_{sun}$ ranges between 20 and 30 minutes except at large $|\Delta\phi|$, and the path length is 1.9~au or longer at all relative heliolongitudes, 60-130\% longer than that for weak turbulence. The SEPs have access to wider range of heliolongitudes in the case of strong turbulence.

For very strong turbulence ($dB^2/B^2=2$ at 1~au, red circles in Figure~\ref{fig:pathlen_time}), the  solar injection times obtained from VDA are considerably late, from tens to hundreds of minutes after the injection in the simulations. The apparent path length varies from 5 to 20~au,  significantly longer than typical VDA results \citep[e.g.][]{Vainio2013_firstSEPServerevent, Paassilta2017_Catalogue55-80MeV} and may indicate that the turbulence parameters used in our simulations for the very strong turbulence are not realistic. Indeed, our results suggest that the VDA analysis record can be used as a constraint for the interplanetary turbulence parameters. We note that only the relative turbulence amplitude is varied in this work: the turbulence composition, turbulence length scales, as well as their variation in the heliosphere are also likely to affect SEP transport, thus the conclusion on the very strong turbulence presented here is valid only for the particular setup adopted in this study. Varying other parameters will be left to future work.

\begin{figure}
    \centering
    \includegraphics[width=\columnwidth]{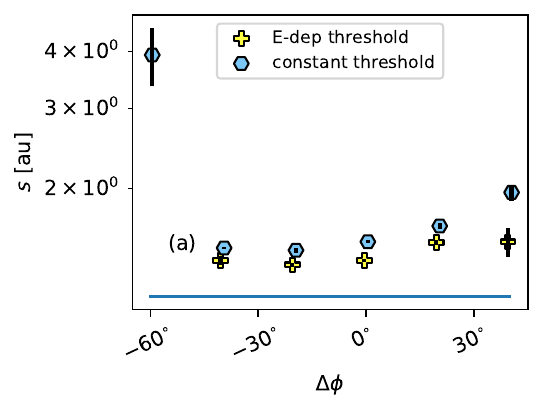}
    \includegraphics[width=\columnwidth]{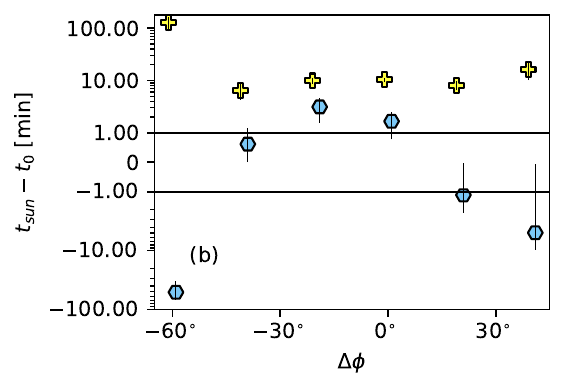}
    \caption{The VDA apparent path length (a) and solar injection time (b) of SEPs as a function of the heliolongitude for energy-dependent (Equation~(\ref{eq:thresh_E}), yellow cross) and constant (blue hexagon) threshold fraction $f(E)=0.01$, for moderate turbulence. In panel (a), the apparent path length $s=-0.05$~au for $\Delta\phi=-60^\circ$ is omitted for clarity.}\label{fig:threshspect_comp}
  \end{figure}

The VDA results shown in Figure~\ref{fig:pathlen_time} are derived using an energy-dependent threshold fraction, Equation~(\ref{eq:thresh_E}), justified by most SEP events reaching higher intensities above the pre-event background at lower energies. However, the relation between the background spectrum and the SEP event peak spectrum can vary considerably, for example due to previous SEP events. The effect of such difference in background and SEP event spectra on VDA results was highlighted in \citet{Laitinen2015_CorrectingInterplanetaryScattering}, who used 1D SEP simulations. We investigate the effect of varying the assumed pre-event spectrum in Figure~\ref{fig:threshspect_comp}, where we compare VDA results for our threshold fraction, Equation~(\ref{eq:thresh_E}) (yellow crosses), and an energy-independent threshold fraction with $f(E)=0.01$ (cyan hexagons).
Note that we have omitted the apparent path length $s=-0.05$~au for $\Delta\phi=-60^\circ$ for clarity: this value was derived with a small number of energy channels that could be selected for VDA using our criteria, and low statistics.
We find that with the constant threshold fraction, $s$ is about 5-10\% longer for $\Delta\phi\in[-40^\circ,40^\circ]$, whereas the difference in $t_{sun}$ varies between 5 and 20~minutes, with the constant threshold fraction, corresponding to equal background and SEP event spectra, resulting in earlier $t_{sun}$. 

This is a direct consequence of the influence of the threshold, or pre-event background in observed SEP events, on the derived onset time at each energy channel. When the threshold is increased or decreased, the derived onset time moves to later or earlier time, respectively. For the constant threshold fraction case, the low-energy (high-$\beta^{-1}$) channel onset times are moved to later and high-energy (low-$\beta^{-1}$) ones to earlier. It is easy to see for example by viewing Figure~\ref{fig:veldisp} that this results in reduction of the slope of the velocity dispersion pattern, which in turn results in longer VDA path length and earlier solar injection time. 

In this work, we have concentrated on investigating the effects of the turbulence strength, relative heliolongitude $\Delta\phi$ and onset threshold on the VDA results. In doing so, we chose a single parametrisation of the SEP injection function, Equation~(\ref{eq:inj}). The effect of changes in the temporal injection profile have already been investigate by \citet{Laitinen2010_EffectPre-eventBackground}, using a 1D SEP transport model. Their findings show that a slow injection results typically in a VDA path length considerably longer than the nominal Parker spiral length. A wider SEP source would likely result in reduced dependence of the VDA results on $\Delta\phi$.

We note that adiabatic deceleration, omitted in our current model, will also affect the VDA analysis. While we will address this issue fully in a future work, we can analyse the effect through the momentum loss equation 
\begin{equation}\label{eq:decel}
\pder{p}{t} = -\frac{p}{3}\nabla\cdot \mathbf{v}_{sw},    
\end{equation}
valid for an isotropic particle distribution \citep[e.g.][]{Ruffolo1995_EffectAdiabaticDeceleration}. For spherically-symmetric solar wind with constant speed, we can derive a deceleration timescale $\tau_d=\frac{3r}{2 v_{sw}}$ which is independent of the particle momentum. As a simple estimate we evaluate $\tau_d$ at 0.5~au, with $v_{sw}=400 \mathrm{\,km/s}$, to calculate how much the SEPs decelerate during their propagation. Comparing the resulting arrival times of scatter-free SEPs at 1~au shows that a 1~MeV proton is delayed by $\sim 4\%$, whereas a 100-MeV proton is delayed by $<1\%$, resulting in lengthening of the path length of $\sim 4\%$. It should be noted that this estimate depends on significant simplifications, such as assumption of isotropic distribution of the first SEPs. We will return to the issue in a future study.

Our results indicate that VDA-derived parameters are influenced by turbulence even when it is weak. Even in this case, propagation through the turbulent medium results in delayed arrival of SEPs. Within the diffusion approximation the onset delays were recently investigated by \citet{Strauss2023_OnsetDelaysSolar}. The delays associated with meandering of magnetic field lines have also received recent interest \citep{Laitinen2019_FromSunto, Chhiber2021_Magneticfieldline, Laitinen2023_AnalyticalModelTurbulence}.  These studies have found a significant dependence of delays on $\Delta\phi$. Our analysis shows that an uncertainty of order 10-20 minutes for $t_{sun}$ for 1-100~MeV protons may result from the interplanetary turbulence and the pre-event background effects. We further point out that overall increase of the pre-event background levels is likely to worsen the VDA accuracy \citep[see, e.g.][]{Lintunen2004_Solarenergeticparticle, Saiz2005_EstimationSolarEnergetic, Laitinen2015_CorrectingInterplanetaryScattering}.

Finally, we note that our results use a particular set of turbulence parameters, the justification for which is outlined in \citet{Laitinen2023_AnalyticalModelTurbulence} 
, varying only the turbulence amplitude. The effect of choosing a different set of turbulence parameters, including the slab/2D power ratio, spectral indices, correlation scales, and their radial evolution, is likely to have complex consequences to the field line meandering and SEP scattering in the heliosphere, and evaluation of the effects requires a full parameter study.
In near future we aim to extend our study to wider range of turbulence parameters, informed by the recent theoretical \citep[e.g.][]{Zhao2017_CosmicRayDiffusiona, Zank2017_TheoryTransportNearly, Zank2018_TheoryTransportNearly, Usmanov2025_UnifiedThreedimensionalMagnetohydrodynamic} and observational \citep[e.g.][]{Zhao2018_InfluenceSolarCycle,Telloni2021_EvolutionSolarWind, Chhiber2022_AnisotropicMagneticTurbulence,Zhao2022_TurbulentPropertiesSub-Alfvenic, Zhu2025_RadialEvolutionMHD} research.

\section{Conclusions}\label{sec:conclusions}

In this work, we have used full-orbit test particle simulations in our recent new model of heliospheric turbulence in Parker spiral geometry to investigate the often used VDA method for SEP event onset analysis. In our study, both the initial SEP propagation along meandering field lines, and the Parker spiral geometry, are taken into account. We simulated the propagation of $1-100$ MeV protons and obtained synthetic time-intensity profiles at heliolongitudinally-distributed fictional observers. We then derived onset times $t_{onset}$ using two different onset threshold definitions and used Equation~(\ref{eq:VDA}) to obtain the solar injection time $t_{sun}$ and the apparent path length $s$ of the protons, for four cases with different relative turbulence amplitude. The levels of turbulence considered correspond to the so-called Palmer consensus range of parallel scattering mean free paths \citep{Palmer1982_Transportcoefficientslow-energy}.

We find that
\begin{itemize}
    \item The VDA-derived solar release times are within 30 minutes from the release time of the simulated particles for the weak,  moderate and strong turbulence cases, with $dB^2/B^2=0.2$, $0.6$ and $1$ at 1~au, respectively. We note that the first SEPs are not scatter-free even in the weak-turbulence case. 
    \item The difference in the turbulence variance between the weak and moderate turbulence cases results in 3 to 15 minute $\Delta\phi$-dependent difference in the derived VDA solar injection time, and 0.2-0.3~au difference in the VDA path length. The difference between the weak and strong turbulence cases increases up to 25 minutes and 1.8 au, respectively.
    \item For the very strong turbulence, with $dB^2/B^2=2$ at 1~au, corresponding to the low scattering mean free path limit of the Palmer consensus, produces  $s>5$~au, rarely seen in VDA analysis of SEP events \citep[e.g.][]{Paassilta2017_Catalogue55-80MeV}, and solar release time error of order 100~minutes. This may indicate that the turbulence conditions represented by our very strong simulation case are not realistic.
    \item The details of energy-dependence of the onset threshold, mimicking the effect of the pre-event background in observed SEP events, affect the VDA results significantly, with 5-10\% difference in VDA path lengths, and 5-20-minute difference in VDA solar injection times between energy-dependent and energy-independent onset thresholds.
\end{itemize}

Our results indicate that in weak turbulence conditions the VDA results can be fairly accurate. In moderate to strong turbulence both the path lengthening likely due to meandering of field lines, and scattering of particles even at the start of the SEP event, reduce the accuracy considerably particularly when the observer is not well connected to the source along the nominal Parker spiral. These effects of turbulence are naturally exacerbated by the effects of high and energy-dependent pre-event cosmic ray background, which can affect both the VDA solar release timing and the apparent path length \citep[and results presented here]{Lintunen2004_Solarenergeticparticle, Saiz2005_EstimationSolarEnergetic, Laitinen2015_CorrectingInterplanetaryScattering}\, as well as instrumental effects \citep[e.g.][]{Li2013_ElectronProtonAcceleration}. Thus, in many situations VDA is not an accurate method for estimating the SEP path lengths and the solar injection times. Specifically VDA is inaccurate  when the turbulence levels are not particularly low and when the observer is not well magnetically connected. We finally note that the level and the spectrum of the pre-event background can significantly increase the uncertainty of the VDA results.

Future work will investigate the turbulence parameter space in more detail, and take into account the effect of adiabatic deceleration, not included in the present study: this is expected to delay the arrival of low energy SEPs in the simulations and produce a small increase (estimated to be less than 4\%) in derived path lengths.

\begin{acknowledgements}
TL and SD acknowledge support from the UK Science and Technology Facilities Council (STFC) through grants ST/V000934/1 and ST/Y002725/1. For the purpose of open access, the author has applied a Creative Commons Attribution (CC BY) licence to any Author’s Accepted Manuscript version arising from this submission. This research was supported by the International Space Science Institute (ISSI) in Bern, through ISSI International Team project \#24-608 (Energetic Particle Transport in Space Plasma Turbulence). The simulations were run with the High-performance computing facility at the Jeremiah Horrocks Institute at University of Lancashire.
\end{acknowledgements}

\section*{Data Availability}

Time-intensity profiles used in this study are available as netCDF
files via Zenodo at 
\dataset[doi:10.5281/zenodo.22028341]{https://doi.org/10.5281/zenodo.22028341}. Data used to produce Figures 3 and 4 are also provided.

\bibliographystyle{aasjournal}
\bibliography{ms}

\begin{thebibliography}{}
\expandafter\ifx\csname natexlab\endcsname\relax\def\natexlab#1{#1}\fi
\providecommand{\url}[1]{\href{#1}{#1}}
\providecommand{\dodoi}[1]{doi:~\href{http://doi.org/#1}{\nolinkurl{#1}}}
\providecommand{\doeprint}[1]{\href{http://ascl.net/#1}{\nolinkurl{http://ascl.net/#1}}}
\providecommand{\doarXiv}[1]{\href{https://arxiv.org/abs/#1}{\nolinkurl{https://arxiv.org/abs/#1}}}

\bibitem[{Adhikari {et~al.}(2024)Adhikari, Zank, Zhao, Wang, Tang, Telloni,
  Pitna, \& Nykyri}]{Adhikari2024_MHDInertialEnergycontaining}
Adhikari, L., Zank, G.~P., Zhao, L., {et~al.} 2024, The Astrophysical Journal,
  965, 94, \dodoi{10.3847/1538-4357/ad2fc4}

\bibitem[{Adhikari {et~al.}(2022)Adhikari, Zank, Zhao, \&
  Telloni}]{Adhikari2022_2DSlabTurbulent}
Adhikari, L., Zank, G.~P., Zhao, L.~L., \& Telloni, D. 2022, 938, 120,
  \dodoi{10.3847/1538-4357/ac9234}

\bibitem[{{Bieber} {et~al.}(1996){Bieber}, {Wanner}, \&
  {Matthaeus}}]{Bieber1996_Dominanttwo-dimensionalsolar}
{Bieber}, J.~W., {Wanner}, W., \& {Matthaeus}, W.~H. 1996, \jgr, 101, 2511,
  \dodoi{10.1029/95JA02588}

\bibitem[{Chhiber(2022)}]{Chhiber2022_AnisotropicMagneticTurbulence}
Chhiber, R. 2022, 939, 33, \dodoi{10.3847/1538-4357/ac9386}

\bibitem[{{Chhiber} {et~al.}(2021){Chhiber}, {Matthaeus}, {Cohen}, {Ruffolo},
  {Sonsrettee}, {Tooprakai}, {Seripienlert}, {Chuychai}, {Usmanov},
  {Goldstein}, {McComas}, {Leske}, {Szalay}, {Joyce}, {Cummings}, {Roelof},
  {Christian}, {Mewaldt}, {Labrador}, {Giacalone}, {Schwadron}, {Mitchell},
  {Hill}, {Wiedenbeck}, {McNutt}, \& {Desai}}]{Chhiber2021_Magneticfieldline}
{Chhiber}, R., {Matthaeus}, W.~H., {Cohen}, C.~M.~S., {et~al.} 2021, \aap, 650,
  A26, \dodoi{10.1051/0004-6361/202039816}

\bibitem[{{Dalla} {et~al.}(2003){Dalla}, {Balogh}, {Krucker}, {Posner},
  {M{\"u}ller-Mellin}, {Anglin}, {Hofer}, {Marsden}, {Sanderson}, {Heber},
  {Zhang}, \& {McKibben}}]{Dalla2003_Delaysolarenergetic}
{Dalla}, S., {Balogh}, A., {Krucker}, S., {et~al.} 2003, Annales Geophysicae,
  21, 1367, \dodoi{10.5194/angeo-21-1367-2003}

\bibitem[{Ding {et~al.}(2022)Ding, Wijsen, Li, \&
  Poedts}]{Ding2022_Modeling2020November}
Ding, Z., Wijsen, N., Li, G., \& Poedts, S. 2022, Astronomy \& Astrophysics,
  668, A71, \dodoi{10.1051/0004-6361/202244732}

\bibitem[{{Dresing} {et~al.}(2012){Dresing}, {G{\'o}mez-Herrero}, {Klassen},
  {Heber}, {Kartavykh}, \& {Dr{\"o}ge}}]{Dresing2012_LargeLongitudinalSpread}
{Dresing}, N., {G{\'o}mez-Herrero}, R., {Klassen}, A., {et~al.} 2012, \solphys,
  281, 281, \dodoi{10.1007/s11207-012-0049-y}

\bibitem[{Giacalone {et~al.}(2020)Giacalone, Mitchell, Allen, Hill, McNutt,
  Szalay, Desai, Rouillard, Kouloumvakos, McComas, Christian, Schwadron,
  Wiedenbeck, Bale, Brown, Case, Chen, Cohen, Joyce, Kasper, Klein, Korreck,
  Larson, Livi, Leske, MacDowall, Matthaeus, Mewaldt, Nieves-Chinchilla,
  Pulupa, Roelof, Stevens, Szabo, \&
  Whittlesey}]{Giacalone2020_SolarEnergeticParticles}
Giacalone, J., Mitchell, D.~G., Allen, R.~C., {et~al.} 2020, The Astrophysical
  Journal Supplement Series, 246, 29, \dodoi{10.3847/1538-4365/ab5221}

\bibitem[{{Huttunen-Heikinmaa} {et~al.}(2005){Huttunen-Heikinmaa}, {Valtonen},
  \& {Laitinen}}]{Huttunen-Heikinmaa2005_Protonheliumrelease}
{Huttunen-Heikinmaa}, K., {Valtonen}, E., \& {Laitinen}, T. 2005, \aap, 442,
  673, \dodoi{10.1051/0004-6361:20042620}

\bibitem[{{Jokipii}(1966)}]{Jokipii1966_Cosmic-RayPropagationI}
{Jokipii}, J.~R. 1966, \apj, 146, 480, \dodoi{10.1086/148912}

\bibitem[{{Krucker} \& {Lin}(2000)}]{Krucker2000_TwoClassesSolar}
{Krucker}, S., \& {Lin}, R.~P. 2000, \apjl, 542, L61, \dodoi{10.1086/312922}

\bibitem[{{Laitinen} \& {Dalla}(2019)}]{Laitinen2019_FromSunto}
{Laitinen}, T., \& {Dalla}, S. 2019, \apj, 887, 222,
  \dodoi{10.3847/1538-4357/ab54c7}

\bibitem[{Laitinen \& Dalla(2025)}]{Laitinen2025_InterplayLargescaleDrift}
Laitinen, T., \& Dalla, S. 2025, The Astrophysical Journal, 979, 106,
  \dodoi{10.3847/1538-4357/ad9c3b}

\bibitem[{{Laitinen} {et~al.}(2023{\natexlab{a}}){Laitinen}, {Dalla},
  {Waterfall}, \& {Hutchinson}}]{Laitinen2023_AnalyticalModelTurbulence}
{Laitinen}, T., {Dalla}, S., {Waterfall}, C.~O.~G., \& {Hutchinson}, A.
  2023{\natexlab{a}}, \apj, 943, 108, \dodoi{10.3847/1538-4357/aca892}

\bibitem[{{Laitinen} {et~al.}(2023{\natexlab{b}}){Laitinen}, {Dalla},
  {Waterfall}, \& {Hutchinson}}]{Laitinen2023_Solarenergeticparticle}
---. 2023{\natexlab{b}}, \aap, 673, L8, \dodoi{10.1051/0004-6361/202346384}

\bibitem[{{Laitinen} {et~al.}(2010){Laitinen}, {Huttunen-Heikinmaa}, \&
  {Valtonen}}]{Laitinen2010_EffectPre-eventBackground}
{Laitinen}, T., {Huttunen-Heikinmaa}, K., \& {Valtonen}, E. 2010, in American
  Institute of Physics Conference Series, Vol. 1216, Twelfth International
  Solar Wind Conference, ed. M.~{Maksimovic}, K.~{Issautier},
  N.~{Meyer-Vernet}, M.~{Moncuquet}, \& F.~{Pantellini}, 249--252,
  \dodoi{10.1063/1.3395847}

\bibitem[{{Laitinen} {et~al.}(2015){Laitinen}, {Huttunen-Heikinmaa},
  {Valtonen}, \& {Dalla}}]{Laitinen2015_CorrectingInterplanetaryScattering}
{Laitinen}, T., {Huttunen-Heikinmaa}, K., {Valtonen}, E., \& {Dalla}, S. 2015,
  \apj, 806, 114, \dodoi{10.1088/0004-637X/806/1/114}

\bibitem[{Li {et~al.}(2013)Li, Firoz, Sun, \&
  Miroshnichenko}]{Li2013_ElectronProtonAcceleration}
Li, C., Firoz, K.~A., Sun, L.~P., \& Miroshnichenko, L.~I. 2013, The
  Astrophysical Journal, 770, 34, \dodoi{10.1088/0004-637X/770/1/34}

\bibitem[{{Li} \& {Bian}(2023)}]{Li2023_LagrangianStochasticModel}
{Li}, G., \& {Bian}, N.~H. 2023, \apj, 945, 150,
  \dodoi{10.3847/1538-4357/acbd43}

\bibitem[{{Lin} {et~al.}(1981){Lin}, {Potter}, {Gurnett}, \&
  {Scarf}}]{Lin1981_Energeticelectronsplasma}
{Lin}, R.~P., {Potter}, D.~W., {Gurnett}, D.~A., \& {Scarf}, F.~L. 1981, \apj,
  251, 364, \dodoi{10.1086/159471}

\bibitem[{{Lintunen} \& {Vainio}(2004)}]{Lintunen2004_Solarenergeticparticle}
{Lintunen}, J., \& {Vainio}, R. 2004, \aap, 420, 343,
  \dodoi{10.1051/0004-6361:20034247}

\bibitem[{Logachev {et~al.}(2002)Logachev, Kecskeméty, \&
  Zeldovich}]{Logachev2002_EnergySpectraLowFlux}
Logachev, Y.~I., Kecskeméty, K., \& Zeldovich, M.~A. 2002, Solar Physics, 208,
  141, \dodoi{10.1023/A:1019689101515}

\bibitem[{{Marsh} {et~al.}(2013){Marsh}, {Dalla}, {Kelly}, \&
  {Laitinen}}]{Marsh2013_Drift-inducedPerpendicularTransport}
{Marsh}, M.~S., {Dalla}, S., {Kelly}, J., \& {Laitinen}, T. 2013, \apj, 774, 4,
  \dodoi{10.1088/0004-637X/774/1/4}

\bibitem[{{Matthaeus} {et~al.}(1990){Matthaeus}, {Goldstein}, \&
  {Roberts}}]{Matthaeus1990_Evidencepresencequasi-two-dimensional}
{Matthaeus}, W.~H., {Goldstein}, M.~L., \& {Roberts}, D.~A. 1990, \jgr, 95,
  20673, \dodoi{10.1029/JA095iA12p20673}

\bibitem[{{Matthaeus} {et~al.}(2003){Matthaeus}, {Qin}, {Bieber}, \&
  {Zank}}]{Matthaeus2003_NonlinearCollisionlessPerpendicular}
{Matthaeus}, W.~H., {Qin}, G., {Bieber}, J.~W., \& {Zank}, G.~P. 2003, \apjl,
  590, L53, \dodoi{10.1086/376613}

\bibitem[{Mewaldt {et~al.}(2005)Mewaldt, Cohen, Labrador, Leske, Mason, Desai,
  Looper, Mazur, Selesnick, \& Haggerty}]{Mewaldt2005_ProtonHeliumElectrona}
Mewaldt, R.~A., Cohen, C. M.~S., Labrador, A.~W., {et~al.} 2005, Journal of
  Geophysical Research (Space Physics), 110, A09S18,
  \dodoi{10.1029/2005JA011038}

\bibitem[{Mewaldt {et~al.}(2012)Mewaldt, Looper, Cohen, Haggerty, Labrador,
  Leske, Mason, Mazur, \& von
  Rosenvinge}]{Mewaldt2012_EnergySpectraCompositiona}
Mewaldt, R.~A., Looper, M.~D., Cohen, C. M.~S., {et~al.} 2012, Space Science
  Reviews, 171, 97, \dodoi{10.1007/s11214-012-9884-2}

\bibitem[{{Moradi} \& {Li}(2019)}]{Moradi2019_PropagationScatter-freeSolar}
{Moradi}, A., \& {Li}, G. 2019, \apj, 887, 102,
  \dodoi{10.3847/1538-4357/ab4f68}

\bibitem[{{Oughton} {et~al.}(2015){Oughton}, {Matthaeus}, {Wan}, \&
  {Osman}}]{Oughton2015_Anisotropysolarwind}
{Oughton}, S., {Matthaeus}, W.~H., {Wan}, M., \& {Osman}, K.~T. 2015,
  Philosophical Transactions of the Royal Society of London Series A, 373,
  20140152, \dodoi{10.1098/rsta.2014.0152}

\bibitem[{{Paassilta} {et~al.}(2018){Paassilta}, {Papaioannou}, {Dresing},
  {Vainio}, {Valtonen}, \& {Heber}}]{Paassilta2018_Catalogue55MeV}
{Paassilta}, M., {Papaioannou}, A., {Dresing}, N., {et~al.} 2018, \solphys,
  293, 70, \dodoi{10.1007/s11207-018-1284-7}

\bibitem[{{Paassilta} {et~al.}(2017){Paassilta}, {Raukunen}, {Vainio},
  {Valtonen}, {Papaioannou}, {Siipola}, {Riihonen}, {Dierckxsens}, {Crosby},
  {Malandraki}, {Heber}, \& {Klein}}]{Paassilta2017_Catalogue55-80MeV}
{Paassilta}, M., {Raukunen}, O., {Vainio}, R., {et~al.} 2017, Journal of Space
  Weather and Space Climate, 7, A14, \dodoi{10.1051/swsc/2017013}

\bibitem[{{Palmer}(1982)}]{Palmer1982_Transportcoefficientslow-energy}
{Palmer}, I.~D. 1982, Reviews of Geophysics and Space Physics, 20, 335,
  \dodoi{10.1029/RG020i002p00335}

\bibitem[{Palmroos {et~al.}(2025)Palmroos, Dresing, Gieseler, Gutiérrez, \&
  Vainio}]{Palmroos2025_NewMethodDetermining}
Palmroos, C., Dresing, N., Gieseler, J., Gutiérrez, C.~P., \& Vainio, R. 2025,
  Astronomy and Astrophysics, 694, A221, \dodoi{10.1051/0004-6361/202451280}

\bibitem[{{Parker}(1958)}]{Parker1958_DynamicsInterplanetaryGas}
{Parker}, E.~N. 1958, \apj, 128, 664, \dodoi{10.1086/146579}

\bibitem[{{Pei} {et~al.}(2006){Pei}, {Jokipii}, \&
  {Giacalone}}]{Pei2006_EffectRandomMagnetic}
{Pei}, C., {Jokipii}, J.~R., \& {Giacalone}, J. 2006, \apj, 641, 1222,
  \dodoi{10.1086/427161}

\bibitem[{Posner {et~al.}(2024)Posner, Richardson, \&
  Strauss}]{Posner2024_SEPClockDiscussion}
Posner, A., Richardson, I.~G., \& Strauss, R. D.~T. 2024, Solar Physics, 299,
  126, \dodoi{10.1007/s11207-024-02350-7}

\bibitem[{{Reames}(2009)}]{Reames2009_SolarEnergetic-ParticleRelease}
{Reames}, D.~V. 2009, \apj, 706, 844, \dodoi{10.1088/0004-637X/706/1/844}

\bibitem[{{Reames} {et~al.}(1985){Reames}, {von Rosenvinge}, \&
  {Lin}}]{Reames1985_SolarHe-3-richevents}
{Reames}, D.~V., {von Rosenvinge}, T.~T., \& {Lin}, R.~P. 1985, \apj, 292, 716,
  \dodoi{10.1086/163203}

\bibitem[{{Richardson} {et~al.}(2014){Richardson}, {von Rosenvinge}, {Cane},
  {Christian}, {Cohen}, {Labrador}, {Leske}, {Mewaldt}, {Wiedenbeck}, \&
  {Stone}}]{Richardson2014_25MeVProton}
{Richardson}, I.~G., {von Rosenvinge}, T.~T., {Cane}, H.~V., {et~al.} 2014,
  \solphys, 289, 3059, \dodoi{10.1007/s11207-014-0524-8}

\bibitem[{{Ruffolo}(1995)}]{Ruffolo1995_EffectAdiabaticDeceleration}
{Ruffolo}, D. 1995, \apj, 442, 861, \dodoi{10.1086/175489}

\bibitem[{{S{\'a}iz} {et~al.}(2005){S{\'a}iz}, {Evenson}, {Ruffolo}, \&
  {Bieber}}]{Saiz2005_EstimationSolarEnergetic}
{S{\'a}iz}, A., {Evenson}, P., {Ruffolo}, D., \& {Bieber}, J.~W. 2005, \apj,
  626, 1131, \dodoi{10.1086/430293}

\bibitem[{Shalchi(2009)}]{Shalchi2009_NonlinearCosmicRay}
Shalchi, A. 2009, Nonlinear {Cosmic} {Ray} {Diffusion} {Theories}, Vol. 362,
  \dodoi{10.1007/978-3-642-00309-7}

\bibitem[{Strauss {et~al.}(2023)Strauss, Dresing, Richardson, van~den Berg, \&
  Steyn}]{Strauss2023_OnsetDelaysSolar}
Strauss, R.~D., Dresing, N., Richardson, I.~G., van~den Berg, J.~P., \& Steyn,
  P.~J. 2023, The Astrophysical Journal, 951, 2,
  \dodoi{10.3847/1538-4357/acd3ef}

\bibitem[{Telloni {et~al.}(2021)Telloni, Sorriso-Valvo, Woodham, Panasenco,
  Velli, Carbone, Zank, Bruno, Perrone, Nakanotani, Shi, D'Amicis, Marco,
  Jagarlamudi, Steinvall, Marino, Adhikari, Zhao, Liang, Tenerani, Laker,
  Horbury, Bale, Pulupa, Malaspina, MacDowall, Goetz, Wit, Harvey, Kasper,
  Korreck, Larson, Case, Stevens, Whittlesey, Livi, Owen, Livi, Louarn,
  Antonucci, Romoli, O'Brien, Evans, \&
  Angelini}]{Telloni2021_EvolutionSolarWind}
Telloni, D., Sorriso-Valvo, L., Woodham, L.~D., {et~al.} 2021, The
  Astrophysical Journal Letters, 912, L21, \dodoi{10.3847/2041-8213/abf7d1}

\bibitem[{{Torsti} {et~al.}(1998){Torsti}, {Anttila}, {Kocharov},
  {M{\"a}kel{\"a}}, {Riihonen}, {Sahla}, {Teittinen}, {Valtonen}, {Laitinen},
  \& {Vainio}}]{Torsti1998_Energetic1to}
{Torsti}, J., {Anttila}, A., {Kocharov}, L., {et~al.} 1998, \grl, 25, 2525,
  \dodoi{10.1029/98GL50062}

\bibitem[{{Tylka} {et~al.}(2003){Tylka}, {Cohen}, {Dietrich}, {Krucker},
  {McGuire}, {Mewaldt}, {Ng}, {Reames}, \&
  {Share}}]{Tylka2003_OnsetsReleaseTimes}
{Tylka}, A.~J., {Cohen}, C.~M.~S., {Dietrich}, W.~F., {et~al.} 2003, in
  International Cosmic Ray Conference, Vol.~6, International Cosmic Ray
  Conference, 3305

\bibitem[{Usmanov {et~al.}(2025)Usmanov, Chhiber, Matthaeus, Roy, \&
  Goldstein}]{Usmanov2025_UnifiedThreedimensionalMagnetohydrodynamic}
Usmanov, A.~V., Chhiber, R., Matthaeus, W.~H., Roy, S., \& Goldstein, M.~L.
  2025, The Astrophysical Journal, 993, 87, \dodoi{10.3847/1538-4357/ae019c}

\bibitem[{{Vainio} {et~al.}(2013){Vainio}, {Valtonen}, {Heber}, {Malandraki},
  {Papaioannou}, {Klein}, {Afanasiev}, {Agueda}, {Aurass}, {Battarbee},
  {Braune}, {Dr{\"o}ge}, {Ganse}, {Hamadache}, {Heynderickx},
  {Huttunen-Heikinmaa}, {Kiener}, {Kilian}, {Kopp}, {Kouloumvakos}, {Maisala},
  {Mishev}, {Miteva}, {Nindos}, {Oittinen}, {Raukunen}, {Riihonen},
  {Rodr{\'\i}guez-Gas{\'e}n}, {Saloniemi}, {Sanahuja}, {Scherer}, {Spanier},
  {Tatischeff}, {Tziotziou}, {Usoskin}, \&
  {Vilmer}}]{Vainio2013_firstSEPServerevent}
{Vainio}, R., {Valtonen}, E., {Heber}, B., {et~al.} 2013, Journal of Space
  Weather and Space Climate, 3, A12, \dodoi{10.1051/swsc/2013030}

\bibitem[{{van den Berg} {et~al.}(2020){van den Berg}, {Strauss}, \&
  {Effenberger}}]{vandenBerg2020_PrimerFocusedSolar}
{van den Berg}, J., {Strauss}, D.~T., \& {Effenberger}, F. 2020, \ssr, 216,
  146, \dodoi{10.1007/s11214-020-00771-x}

\bibitem[{{Wijsen} {et~al.}(2022){Wijsen}, {Aran}, {Scolini}, {Lario},
  {Afanasiev}, {Vainio}, {Sanahuja}, {Pomoell}, \&
  {Poedts}}]{Wijsen2022_Observation-basedmodellingenergetic}
{Wijsen}, N., {Aran}, A., {Scolini}, C., {et~al.} 2022, \aap, 659, A187,
  \dodoi{10.1051/0004-6361/202142698}

\bibitem[{Zank {et~al.}(2017)Zank, Adhikari, Hunana, Shiota, Bruno, \&
  Telloni}]{Zank2017_TheoryTransportNearly}
Zank, G.~P., Adhikari, L., Hunana, P., {et~al.} 2017, The Astrophysical
  Journal, 835, 147, \dodoi{10.3847/1538-4357/835/2/147}

\bibitem[{Zank {et~al.}(2018)Zank, Adhikari, Hunana, Tiwari, Moore, Shiota,
  Bruno, \& Telloni}]{Zank2018_TheoryTransportNearly}
---. 2018, The Astrophysical Journal, 854, 32, \dodoi{10.3847/1538-4357/aaa763}

\bibitem[{{Zhao} {et~al.}(2019){Zhao}, {Li}, {Zhang}, {Wang}, {Moradi}, \&
  {Effenberger}}]{Zhao2019_StatisticalAnalysisInterplanetary}
{Zhao}, L., {Li}, G., {Zhang}, M., {et~al.} 2019, \apj, 878, 107,
  \dodoi{10.3847/1538-4357/ab2041}

\bibitem[{Zhao {et~al.}(2017)Zhao, Adhikari, Zank, Hu, \&
  Feng}]{Zhao2017_CosmicRayDiffusiona}
Zhao, L.-L., Adhikari, L., Zank, G.~P., Hu, Q., \& Feng, X.~S. 2017, The
  Astrophysical Journal, 849, 88, \dodoi{10.3847/1538-4357/aa932a}

\bibitem[{Zhao {et~al.}(2018)Zhao, Adhikari, Zank, Hu, \&
  Feng}]{Zhao2018_InfluenceSolarCycle}
---. 2018, The Astrophysical Journal, 856, 94, \dodoi{10.3847/1538-4357/aab362}

\bibitem[{{Zhao} {et~al.}(2022){Zhao}, {Zank}, {Telloni}, {Stevens}, {Kasper},
  \& {Bale}}]{Zhao2022_TurbulentPropertiesSub-Alfvenic}
{Zhao}, L.~L., {Zank}, G.~P., {Telloni}, D., {et~al.} 2022, \apjl, 928, L15,
  \dodoi{10.3847/2041-8213/ac5fb0}

\bibitem[{Zhu {et~al.}(2025)Zhu, Zank, Zhao, \&
  Silwal}]{Zhu2025_RadialEvolutionMHD}
Zhu, X., Zank, G.~P., Zhao, L., \& Silwal, A. 2025, The Astrophysical Journal,
  978, L34, \dodoi{10.3847/2041-8213/ada354}

\end{thebibliography}

\end{document}